\documentclass[aps,prb,twocolumn,showpacs,amsmath,amssymb,superscriptaddress,citeautoscript]{revtex4-1}

\usepackage{graphicx}
\usepackage{natbib}
\usepackage{dcolumn}
\usepackage{bm}
\usepackage{color}

\begin{document}

\title{Pressure tuning the Fermi-surface topology of the Weyl semimetal NbP}

\author{R.~D.~dos Reis}
\affiliation{Max-Planck Institute for Chemical Physics of Solids, N\"{o}thnitzer Str.\ 40, 01187 Dresden}

\author{S.~C.~Wu}
\affiliation{Max-Planck Institute for Chemical Physics of Solids, N\"{o}thnitzer Str.\ 40, 01187 Dresden}

\author{Y.~Sun}
\affiliation{Max-Planck Institute for Chemical Physics of Solids, N\"{o}thnitzer Str.\ 40, 01187 Dresden}

\author{M.~O.~Ajeesh}
\affiliation{Max-Planck Institute for Chemical Physics of Solids, N\"{o}thnitzer Str.\ 40, 01187 Dresden}

\author{C.~Shekhar}
\affiliation{Max-Planck Institute for Chemical Physics of Solids, N\"{o}thnitzer Str.\ 40, 01187 Dresden}

\author{M.~Schmidt}
\affiliation{Max-Planck Institute for Chemical Physics of Solids, N\"{o}thnitzer Str.\ 40, 01187 Dresden}

\author{C.~Felser}
\affiliation{Max-Planck Institute for Chemical Physics of Solids, N\"{o}thnitzer Str.\ 40, 01187 Dresden}

\author{B.~Yan}
\affiliation{Max-Planck Institute for Chemical Physics of Solids, N\"{o}thnitzer Str.\ 40, 01187 Dresden}
\affiliation{Max Planck Institute for the Physics of Complex Systems, N\"{o}thnitzer Str.\ 38, 01187 Dresden, Germany}
\affiliation{School of Physical Science and Technology, ShanghaiTech University, Shanghai 200031, China}

\author{M.~Nicklas}
\email{michael.nicklas@cpfs.mpg.de}
\affiliation{Max-Planck Institute for Chemical Physics of Solids, N\"{o}thnitzer Str.\ 40, 01187 Dresden}

\begin{abstract}
We report on the pressure evolution of the Fermi surface topology of the Weyl semimetal NbP, probed by Shubnikov-de Haas oscillations in the magnetoresistance combined with \textit{ab-initio} calculations of the band-structure. Although we observe a drastic effect on the amplitudes of the quantum oscillations, the frequencies only exhibit a weak  pressure dependence up to 2.8~GPa. The pressure-induce variations in the oscillation frequencies are consistent with our band-structure  calculations. Furthermore, we can relate the changes in the amplitudes
to small modifications in the shape of the Fermi surface. Our findings evidenced the stability of the electronic band structure of NbP and demonstrate the power of combining quantum-oscillation studies and band-structure calculations to investigate pressure effects on the Fermi-surface topology in Weyl semimetals.

\end{abstract}

\date{\today}

\maketitle

\section{INTRODUCTION}
Recently the Weyl semimetals (WSMs) \cite{Wan2011}  attract enormous attention as a new family of topological materials,\cite{qi2011RMP,Hasan:2010ku,Yan2012rpp} which was stimulated by the recent discovery in  transition-metal monopnictides.\cite{Weng2015,Huang2015,Xu2015TaAs,Lv2015TaAs,Yang2015TaAs} The WSM exhibits topological surface states \cite{Wan2011} that are characterized by Fermi arcs and exotic topological chiral transport properties such as the chiral anomaly effect \cite{Adler1969,Bell1969} and a large magnetoresistance (MR).\cite{Shekhar2015}
While angle-resolved photoemission spectroscopy (ARPES)\cite{Lv2015TaAsbulk,Liu2015NbPTaP,Xu2015NbAs,Xu2015TaP,Souma2015} and theoretical calculations\cite{Sun2015arc,Soluyanov2015,Sun2015MoTe2} have been further employed to address the surface states,
magneto-transport measurements have been also extensively performed recently.\cite{Huang2015anomaly,Zhang2015ABJ,Shekhar2015TaP,Wang:2015wm,Yang:2015vz,Du:2015TaP, Shekhar2015,Moll2015,Zhang2015quantum,Yang2011QHE,Turner:2013tf,Hosur:2013eb,Vafek:2014hl,Parameswaran2014,Baum2015}

In the band structure of a WSM, conduction and valence bands cross each other at nodal points, called Weyl points, and disperse linearly through these Weyl points in the three-dimensional (3D) $k$-space.\cite{Turner:2013tf,Hosur:2013eb} Thus, a WSM can be viewed as a 3D analogue of graphene. The Weyl points appear in pairs, in which the two Weyl points exhibit opposite chiralities. The transition-metal monopnictide WSMs include four members, TaAs, TaP, NbAs, and NbP. These compounds crystalize in a non-centrosymmetric tetragonal lattice structure, but preserve the time-reversal symmetry. There are twelve-pairs of Weyl points that can be classified into two groups: four pairs lie in the $k_z = 0$ plane (labeled as W1) and eight pairs stay in planes of $k_z \sim \pm \pi/c$ (labeled as W2), where $c$ is the lattice constant of the long axis. The 4 compounds exhibit very similar band structures, in which the W1-type Weyl points are slightly lower in energy than the W2-type Weyl points. However, all Weyl points stay slightly away from the Fermi energy with the coexistence of many trivial electron and hole pockets in the Fermi surface (FS), as revealed by recent FS reconstructions in TaP \cite{Shekhar2015TaP} and NbP,\cite{Klotz2015} for example.

The member with the weakest spin-orbit coupling in above four WSM materials, NbP, was found to present an extremely large MR and an ultrahigh mobility.\cite{Shekhar2015} The large MR originates form the combination of the nearly perfect compensation condition \cite{Ali2014} between electron and hole carriers \cite{Shekhar2015,Klotz2015} and the high mobility which is relevant to the topological band structure.
Weyl electrons will generally coexist with normal electrons, and then small changes of the Fermi energy can in principle modify the topology of the Fermi surface due to the smallness of the carrier density. Applying pressure is known to be a powerful approach to tune the electronic structure of a material, which promises a way to vary the energies of Weyl points. Hence, it is interesting to investigate how the positions of the two sets of Weyl points in NbP shift in energy under pressure, and whether pressure can induce other exotic structures and properties in this topological WSM. Since surface-sensitive probes such as ARPES cannot be used for detecting the topological states under pressure, quantum oscillation studies become an ideal tool to determine the effects of pressure on the Fermi-surface topology here. So far, there have been only a few studies on the effect of pressure on NbAs \cite{Luo2016NbAs,Zhang2015NbAs} and TaAs.\cite{Zhou2015TaAs} Although, in all these studies the electronic structure is shown to be very stable, an accurate determination of the effects of pressure on the Fermi-surface topology of Weyl materials is still missing. Shubnikov-de Haas (SdH) oscillations have been reported, but without direct comparison with band-structure calculations.\cite{Luo2016NbAs}

In this paper, we present a pressure study of the magnetotransport properties of the WSM NbP.
By analyzing SdH oscillations observed in the MR, 4 Fermi-surface pockets are revealed, which correspond to two pairs of electron and hole pockets.\cite{Klotz2015} Although we observe a drastic effect on the amplitudes of the quantum oscillations, the frequencies remain almost unaltered up to 2.8~GPa. With support of band-structure calculations we argue that the observed pressure-induced changes in the quantum oscillations arise from small modifications in the shape of the FS. We further relate the strong variation of the high-field MR for magnetic field $B$ and electrical current $I$ parallel to the crystallographic $a$ direction to a combination of different effects, a change in the balance of electron and hole charge carriers and an alteration of the distance of the W2-type Weyl points to the Fermi level which is evidenced by our band structure calculations. Moreover, our findings demonstrate the stability of the electronic band structure of NbP, and show that quantum oscillation studies in association with band-structure calculations are an effective tool to investigate the evolution of the FS topology in Weyl semimetals under pressure.

\section{METHODS}

High-quality single crystals of NbP were grown via a chemical vapor transport reaction. More details on the sample preparation and characterization can be found in Ref.\ \onlinecite{Shekhar2015}. The electrical-transport experiments were performed on high quality NbP single crystals in magnetic fields up to $B=9$~T in a $^{4}$He cryostat (JANIS) equipped with superconducting magnet to temperatures down to $T=1.4$~K. The electrical resistance was measured in 4-point geometry, where the contacts to the sample were made using silver paint and $25~\mu$m gold wire. We prepared two samples taking crystals from the same batch. In both setups $I$ was applied along the crystallographic $a$-axis and the magnetic field perpendicular to the current in sample $S_{B\parallel c}$ and parallel to it in sample $S_{B\parallel a}$. Hydrostatic pressure was generated using a clamp-type pressure cell utilizing silicon oil as pressure transmitting medium. The pressure inside the cell was determined by measuring the shift of the superconducting critical temperature of a piece of Pb.

The \textit{ab-initio} calculations were performed within the framework of the density-functional theory (DFT), implemented in the Vienna \textit{ab-initio} simulation package.\cite{kresse1996} The core electrons were represented by the projector-augmented-wave potential and the generalized gradient approximation (GGA) \cite{perdew1996} was employed for the exchange-correlation functional. A $k$-grid in the first Brillouin zone with Gaussian smearing ($\sigma=0.05$~eV) was utilized. Instead of using experimental lattice parameters, the Murnaghan's equation of state was used to derive the pressure by fitting the total energy dependence of the volume.\cite{Murnaghan1944} The total energy was extracted by optimizing lattice constants and atomic positions for each volume. Then we interpolated the bulk FS using maximally localized Wannier functions.\cite{Mostofi2008}

\section{RESULTS}

\begin{figure}[t!]
\begin{center}
  \includegraphics[clip,width=0.95\columnwidth]{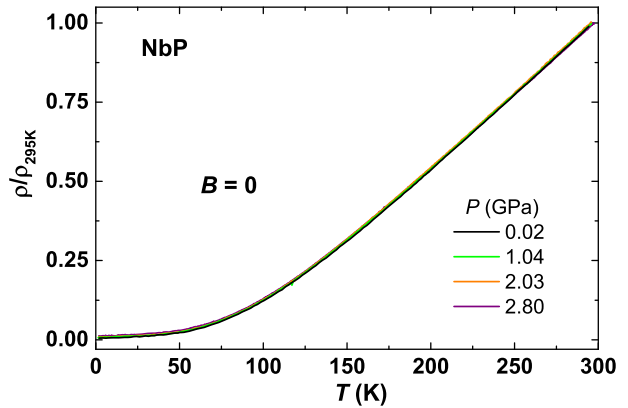}
  \caption{Temperature dependent of the normalized electrical resistivity $\rho/\rho_{\rm295K}$ in zero magnetic field.
  }\label{fig1}
  \end{center}
\end{figure}

The temperature dependence of the longitudinal resistivity $\rho_{xx}(T)$  is depicted in Fig.~\ref{fig1}. At ambient pressure we find a good agreement with previously reported data.\cite{Shekhar2015} Both samples display a metallic behavior with a resistivity ratio $\rho_{xx}({\rm 300\,K})/\rho_{xx}({\rm 2\,K})\approx112$ indicating the high quality of our single crystals. The resistivity ratio is not affected by application of external pressure. The curves obtained at all investigated pressures collapse onto a single line within the experimental resolution. We note that we do not find any indication for superconductivity down to 1.4~K in the whole pressure range $p\leq 2.8$~GPa. At this point, we want to put out that our data are unaltered by pressure even at low temperatures. This is in contrast to observations reported for NbAs,\cite{Luo2016NbAs} where at low temperatures ($T\lesssim30$~K) a small, but monotonous increase in the resistivity upon increasing pressure is eminent in the data. This hints at a more stable electronic structure in NbP compared with NbAs.

\begin{figure}[t!]
\begin{center}
  \includegraphics[clip,width=0.95\columnwidth]{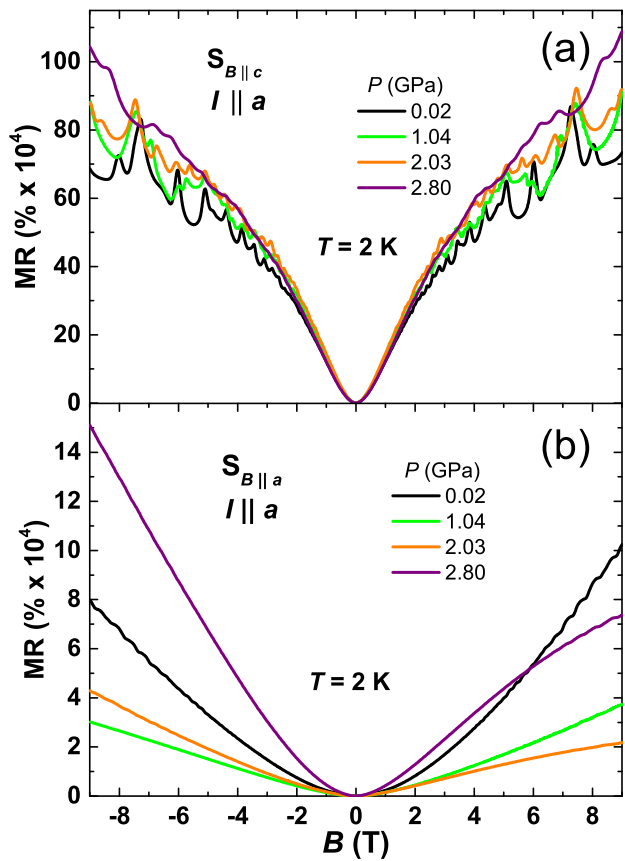}
  \caption{(a) and (b) Magnetoresistance at 2~K for selected pressures in the two configurations $B\parallel c$ and $B\parallel a$, respectively.
  }\label{fig2}
  \end{center}
\end{figure}

\begin{figure}[t!]
\begin{center}
  \includegraphics[clip,width=0.95\columnwidth]{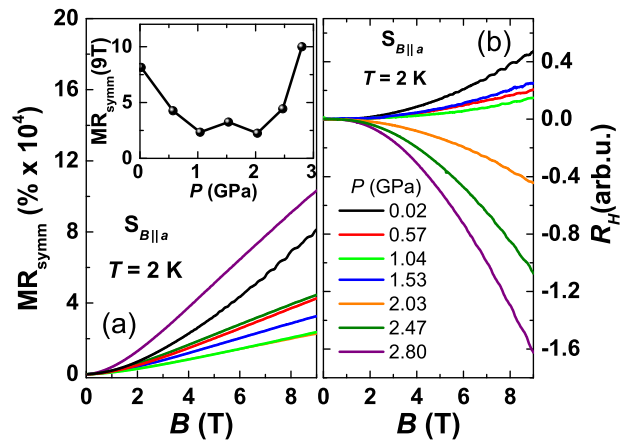}
  \caption{ (a) ${\rm MR_{symm}}$ for $S_{B\parallel a}$ with $B\parallel I$. The inset shows the pressure effects on ${\rm MR_{symm}}$ at 9~T for $S_{B\parallel a}$. (b) Anti-symmetrized MR of sample $S_{B\parallel c}$, which is a measure of the Hall coefficient $R_H$ (see text for details).
  }\label{fig3}
  \end{center}
\end{figure}

In the following we will focus on the MR in NbP. We performed MR measurements under magnetic field up to 9~T and to pressure up to 2.8~GPa. Figures~\ref{fig2}a and~\ref{fig2}b show the MR of samples $S_{B\parallel c}$  and $S_{B\parallel a}$ at 2~K for selected pressures, respectively. Here, we define the MR as the magnetic field dependence of the longitudinal resistivity $\rho_{xx}$: ${\rm MR}(B)=[\rho_{xx}(B)-\rho_{xx}(0)]/{\rho_{xx}(0)}$. Owing to the large charge carrier mobility in NbP,\cite{Shekhar2015} we observe a giant unsaturated MR for both samples. However, the behavior of the MR under pressure is different between the two samples. For sample $S_{B\parallel c}$, the magnitude of the MR remains almost unaltered under pressure, except for oscillatory part, which displays characteristic changes which we will discuss below. On the other hand, for the sample $S_{B\parallel a}$ (see Fig.~\ref{fig2}b) pressure induces dramatic changes in the amplitude of the MR. Furthermore, the MR curves become less symmetric upon increasing pressure. The asymmetry of MR curves originates from a Hall contribution due to a small misalignment of the voltage leads in our experimental setup. An extraordinarily large Hall effect has been previously reported for NbP.\cite{Shekhar2015} The Hall contribution can be removed from the data by a symmetrization of MR curves, ${\rm MR}_{\rm symm}(B)=0.5[{\rm MR}(B)+{\rm MR}(-B)]$. The ${\rm MR}_{\rm symm}$ curves for $S_{B\parallel a}$ are displayed in Fig.~\ref{fig3}a. The pressure dependence of ${\rm MR}_{\rm symm}$ at 9~T (see inset of Fig.~\ref{fig3}a) exhibits a double well structure: upon increasing pressure ${\rm MR}_{\rm symm}({\rm 9T})$ decreases from $8\times10^{4}\%$ at ambient pressure to $2.3\times10^{4}\%$ at 1~GPa, reaches $3.2\times10^{4}\%$ at 1.5~GPa and decreases again to $2\times10^{4}\%$ at 2~GPa. Above 2~GPa ${\rm MR}_{\rm symm}({\rm 9T})$ continues to increase and achieves $10\times10^{4}\%$ at 2.8~GPa which is comparable with the value at 0~GPa. ${\rm MR}_{\rm symm}$ in NbP is directly related to the almost perfect balance between electron- and hole-like charge carriers. The concentration of the charge carriers can be inferred from the Hall coefficient. From our data we can only obtain qualitative information on the Hall effect from the antisymmetric contribution to the MR, ${R_H}(B)\propto0.5[{\rm MR}(B)-{\rm MR}(-B)]$. ${R_H}(B)$ changes sign around 1.5~GPa (see Fig.~\ref{fig3}b). This indicates that the balance between electrons and holes is very sensitive to the application of external pressure. A similar effect has been reported in WTe$_{2}$, which has been identified as a Dirac material.\cite{Cai, Kang2015} In Weyl semimetals the presence of the chiral effect is expected to provide a negative contribution to the longitudinal MR, \textit{i.e.}, when electrical current and magnetic field are parallel.\cite{Adler1969,Bell1969,Nielsen1983} Therefore, a negative contribution to the MR might be expected for sample $S_{B\parallel a}$. However, this is only the case, if the chirality is well-defined, \textit{i.e.}, the Fermi energy is close enough to the Weyl nodes.\cite{Nielsen1983} For NbP at ambient pressure, the two groups of Weyl nodes, W1 and W2, lie -57~meV and 8~meV apart from the Fermi energy, respectively. Thus, a contribution from the chiral anomaly to the MR is unlikely.\cite{Klotz2015} On the other hand, pressure might tune the energy position of the Weyl nodes closer to the Fermi level leading to an increase of the negative contribution from the chiral anomaly to the longitudinal MR. Thus, the pressure variation of the longitudinal MR as observed $S_{B\parallel a}$ could be a combination of different effects.

\begin{figure}[t!]
\begin{center}
  \includegraphics[clip,width=0.9\columnwidth]{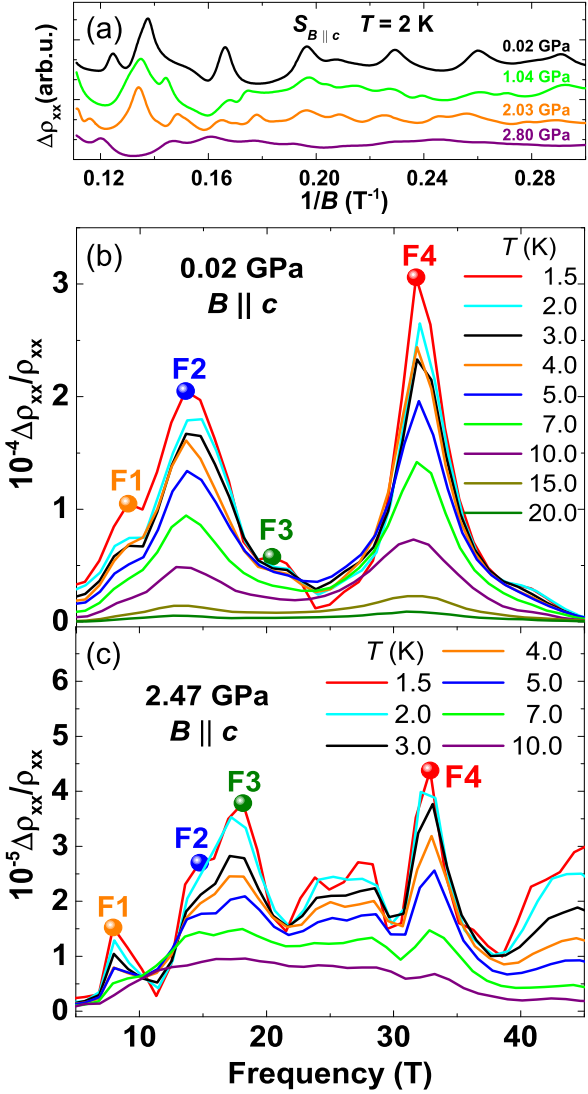}
  \caption{(a) Shubnikov-de Haas oscillations as a function of the inverse magnetic field taken at different pressures for $S_{B\parallel c}$. The oscillatory part was extracted by a subtraction of a third order polynomial background. (b) and (c) temperature dependence of the fast Fourier transform of the SdH oscillations under pressures of 0.02 and 2.47~GPa for $B\parallel c$.}\label{figure4}
  \end{center}
\end{figure}

\begin{figure}[t!]
\begin{center}
  \includegraphics[clip,width=0.9\columnwidth]{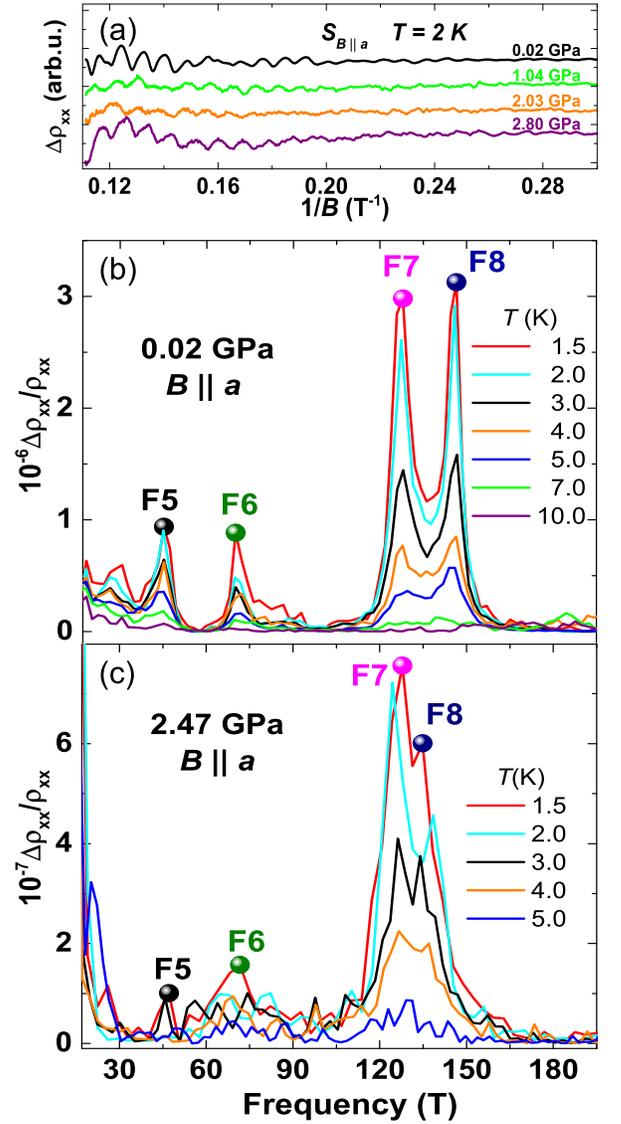}
  \caption{(a) Oscillatory part of MR as a function of the inverse magnetic field taken at different pressures for the sample $S_{B\parallel a}$. (b) and (c) Fast Fourier transform of the SdH oscillations as a function of temperature under pressures of 0.02 and 2.47~GPa for $B\parallel a$.}\label{figure5}
  \end{center}
\end{figure}

The pressure evolution of the Fermi-surface topology can be extracted from the pronounced SdH oscillations in the MR. For both field orientations, $B\parallel c$ and $B\parallel a$, see Figs.~\ref{figure4}a and \ref{figure5}a, respectively, SdH oscillation starting from 1~T are clearly visible, indicating very small effective masses, resulting in high mobilities. The oscillatory part of the signal was obtained by subtracting a third-order polynomial background from the MR data. As expected we find $\Delta\rho_{xx}$ periodic in $1/B$. The SdH frequencies were then determined by a fast Fourier transform on the oscillatory part of the signal.
The 8 fundamental frequencies obtained at ambient pressure, labeled $F1$ to $F8$, are given in Table~I and are in good agreement with literature \cite{Klotz2015}. The corresponding calculated FS pockets including extremal orbits are illustrated in Fig.~\ref{figure6}c and d. In fact, the theory predicts some additional extremal orbits in agreement with our previous work,\cite{Klotz2015} but those could not be resolved in the present experiment. The adopted Fermi level for calculating the extremal orbits is at the ideal electron-hole compensation point.
The frequencies of the oscillations are proportional to extremal Fermi-surface cross-sections $A_{k}$. $A_{k}$ is perpendicular to $B$ following the Onsager relation $F=(\Phi_{0}/2\pi^{2})A_{k}$, where $\Phi_{0}=h/2e$ is the magnetic flux quantum. The temperature dependence of the spectra are displayed in Figs.~\ref{figure4}b and \ref{figure4}c and  in Figs.~\ref{figure5}b and \ref{figure5}c at 0 and 2.47~GPa for $S_{B\parallel c}$ and $S_{B\parallel a}$, respectively. We can identify for both cases 4 fundamental frequencies that remain almost unchanged with pressure. The fundamental frequencies in Figs.~\ref{figure4} and ~\ref{figure5}  are indicated by bullets. The width of the bullets is an estimation of the error in the determination of the frequencies. We point out that beside of the 8 fundamental frequencies, a series of their higher harmonics is also present, even more pronounced in the spectra at higher pressures, manifesting the high sample quality. While the frequencies are almost not affected, application of external pressure dramatically reduces the amplitudes of the quantum-oscillations. The amplitudes of the oscillations are directly related to the curvature $|\partial^{2}A/\partial k_{\parallel}^{2}|$ of the FS cross sections. Therefore, we can argue that the reduction of the amplitude of the quantum oscillations is caused by an increase in the curvature close to an extremal cross section of the FS.

The effective charge-carrier masses $m^*$ of the different orbits are obtained by analyzing the temperature dependencies of the SdH amplitudes. The data for 0 and 2.47~GPa are displayed in Figs.~\ref{figure4} and ~\ref{figure5}. According the Lifshitz-Kosevich formula, the temperature dependence is proportional to $X/\sinh(X)$, with $X=\alpha m^*T/B$ and $\alpha=2\pi^{2}\kappa_{B}m_{e}/(\hbar e)$. Here, $\kappa_{B}$ and $m_{e}$ are the Boltzmann constant and the free-electron mass, respectively. For $B\parallel c$ the masses are rather small, between $m^{*}\approx0.034m_{e}$ and $0.047m_{e}$ at ambient pressure, as expected for semimetals with small FS pockets. In the case of $B\parallel a$ the masses are about twice as large. We find $m^{*}\approx0.104m_{e}$ and $0.112m_{e}$ for the $F7$ and $F8$ frequencies. These results are in good agreement with previous studies.\cite{Klotz2015} The large difference between the effective masses for the two directions is expected due to the highly anisotropic FS pockets in NbP, since in a first approximation $m^{*}$ is proportional to the extremal area of an orbit. Within the experimental resolution, we find almost constant effective masses in the investigated pressure range up to 2.8~GPa. This result further indicates the stability of the electronic structure of NbP. The obtained effective masses at 0, 1.04, and 2.47~GPa are summarized in Table~I.

\begin{table}[t!]
\begin{centering}
\caption{Experimental values of the dHvA frequencies and effective masses for 0, 1.04, and 2.47~GPa  for different orbits. For $F5$ and $F6$ the effective masses could not be determined reliably. The frequency $F$ is in unit of T, and the effective masse $m^{*}$ is in units of free electron masses, $m_{e}$.}
\end{centering}
\label{freq_mass}
\centering{}%
\begin{tabular}{ccclccccccccc}
&&&&&&&&\tabularnewline
\hline
 & & \multicolumn{4}{c}{\textbf{0.02 GPa}} & \multicolumn{3}{c}{\textbf{1.04 GPa}} & & \multicolumn{3}{c}{\textbf{2.47 GPa}}\tabularnewline
 & &  & $F$ & $m^{*}$ &  &  & $F$ & $m^{*}$ &  &  & $F$ & $m^{*}$\tabularnewline
\hline
\hline
 & & \multicolumn{4}{c}{$$} & \multicolumn{3}{c}{$B \parallel c$} & & & \multicolumn{2}{c}{$$}\tabularnewline

$F1$ & $\alpha_2$ &  & ~~9.2 & 0.047(5) &  &  & 10.5 & 0.051(5) &  &  & ~~8.1 & 0.049(6)\tabularnewline
$F2$ & $\beta$ &  & 14.8 & 0.040(6) &  &  & 16.5 & 0.046(5) &  &  & 13.6 & 0.034(5)\tabularnewline
$F3$ & $H2$ &  & 21.8 & 0.034(8) &  &  & 19.1 & 0.038(4) &  &  & 18.9 & 0.044(6)\tabularnewline
$F4$ &       &  & 33.0 & 0.041(7) &  &  & 34.8 & 0.040(5) &  &  & 34.0 & 0.048(4)\tabularnewline
\hline
 & & \multicolumn{4}{c}{$$} & \multicolumn{3}{c}{$B \parallel a$} & & & \multicolumn{2}{c}{$$}\tabularnewline

$F5$ &     &  & 45.1 & - &  &  & 48.6 & - &  &  & 45.35 & -\tabularnewline
$F6$ &     &  & 71.2 & - &  &  & 68.6 & -&  &  & 72.1 & -\tabularnewline
$F7$ &     &  & 128 & 0.112 (10) &  &  & 127.5 & 0.106(8)~~  &  &  & 127.5 & 0.104(8)~~~\tabularnewline
$F8$ &     &  & 146 & 0.104 (12) &  &  & 140.5 & 0.099(10) &  &  & 140.5 & 0.094(13)\tabularnewline
\hline

\end{tabular}

\end{table}

\section{DISCUSSION}

\begin{figure*}[t!]
\begin{center}
  \includegraphics[clip,width=2\columnwidth]{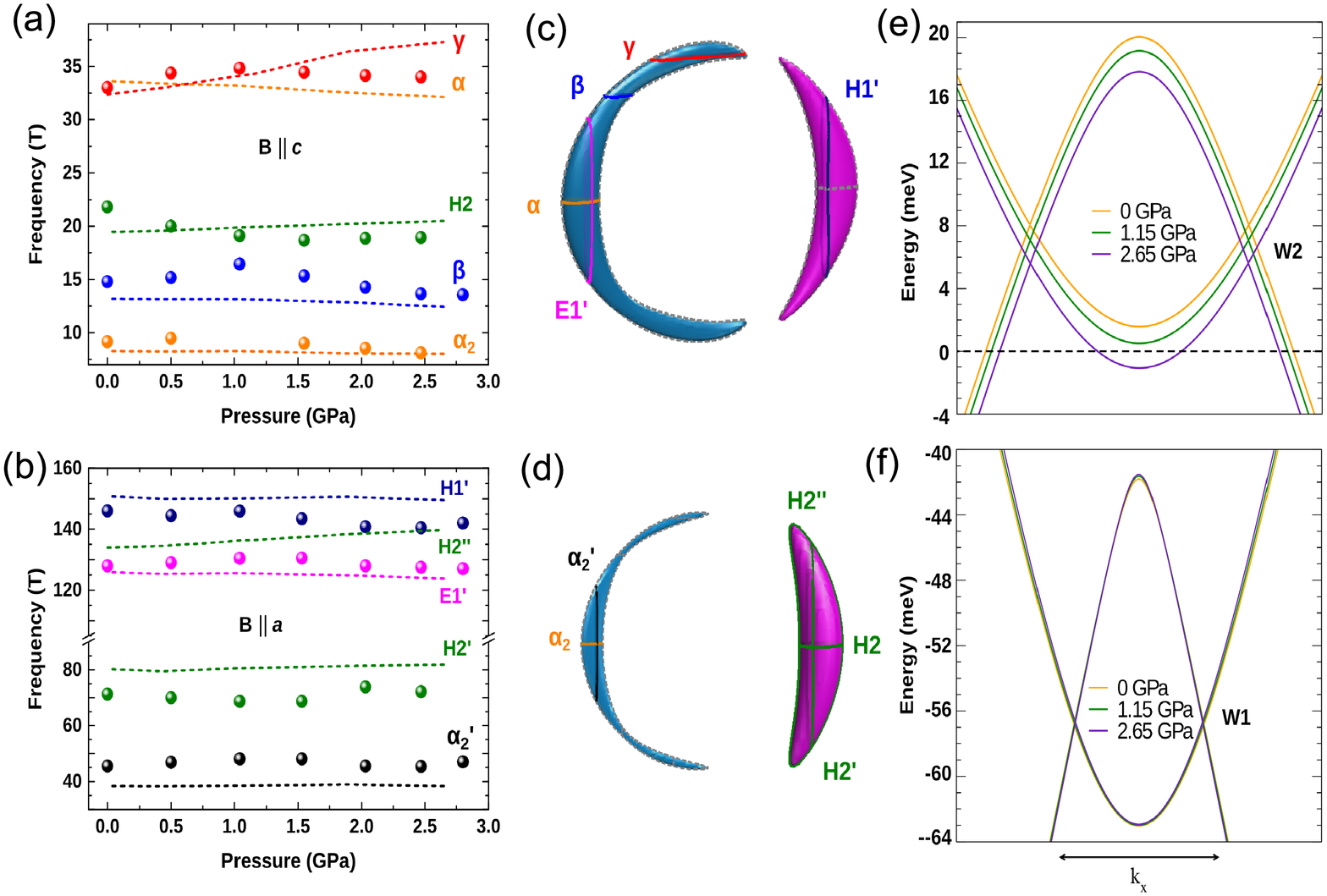}
  \caption{(a) and (b) Pressure dependence of the experimental SdH (solid symbols) and calculated quantum-oscillation frequencies (dashed lines) for $B\parallel c$ and $B\parallel a$, respectively. The size of the bullets gives an estimate of the error in the determination of the oscillation frequencies as shown in Figs.~\ref{figure4} and ~\ref{figure5}. (c) and (d) Illustration of the calculated Fermi-surface pockets [left (blue) - electron and right (violet) - hole pockets]. The extremal orbits identified in our experimental data, $\alpha$, $\alpha_2$, $\alpha_2'$, $\beta$, $\gamma$, $E1'$, $H1'$, $H2$, $H2'$ and $H2''$ are indicated. The gray dash lines mark those extremal orbits that are not confirmed experimentally. Band structures along the line connecting a pair of Weyl points for (e) W2 and (f) W1 at different pressures. The dash line marks the Fermi energy set to zero. }\label{figure6}
  \end{center}
\end{figure*}

In order to get deeper insights in the pressure evolution of the FS topology, we mapped the experimentally obtained frequencies on our calculated FS. The experimental frequencies were taken from the peaks in the spectra obtained from the fast Fourier transforms of the data at 2~K (see Figs.~\ref{figure4},~\ref{figure5} and Table~I). As depicted in the Fig.~\ref{figure6}a and b, we find a good agreement between the pressure dependencies as well as absolute values of the experimental and calculated frequencies. Our calculated FS orbits indicate that the experimental frequencies $F1$, $F2$, $F4$, and $F5$ are related to electron pockets while $F3$ and $F8$ are orbits from hole pockets. Some of the calculated FS orbits are close to one branch of experimental frequencies, such frequencies are experimentally hardly separable. Most likely the $F7$ and $F8$ frequencies are a mixture of $H1'$, $E1'$ and $H2''$ orbits. Sometimes, even though, we know the frequencies belong to one pocket, there are still two different calculated FS orbits close by. Such as the $F4$ frequency branch is composed by $\alpha$ and $\gamma$ orbits of the large electron pocket.
Independent of the field orientation, most of the calculated frequencies belonging to electron orbits ($\alpha$, $\alpha_2$, $\alpha_2'$, $\beta$, and $E1'$) decrease with increasing pressure, except the $\gamma$ orbit. The opposite behavior is observed for the orbits associated with hole pockets ($H1'$, $H2$, $H2'$, and $H2''$) and the arm part of the electron pocket $\gamma$. As we see in our calculations, the effect of pressure is small, which is hard to unravel in the experimental data. Thus, the calculated frequencies can help us to realize the tendency.

Another information we can obtain from our calculations is the energy difference between the Weyl points and the Fermi energy. We picked three pressures and shifted the Fermi energy of the band structures around W1- and W2-type Weyl points to zero. The valley between Weyl points with opposite chiralities in Fig.~\ref{figure6}e and f is almost unchanged, but both Weyl points get more and more closer to the Fermi energy upon increasing pressure. We note that in Fig.~\ref{figure6}e and f the two Weyl points at 0~GPa are not separate by any barrier, which is different compared with our previous work.\cite{Klotz2015} This is caused by carrying out a full relaxation of the lattice parameters in the present work instead of adopting the experimental ones. Here the lattice parameters obtained 2.65~GPa are similar to that used in our previous work.\cite{Klotz2015} Accordingly, we find a tiny barrier of $\sim1$~meV along the line connecting the pair of W2-type Weyl points at 2.65~GPa. However, the Weyl points are still not independent since they merge into the hole pocket in other directions.\cite{Klotz2015} The energy variations of the W1-type Weyl points are small to almost no change (see Fig.~\ref{figure6}f). In contrast to that the W2-type Weyl points exhibit considerable variations. This implies a possibility to induce the chiral anomaly effect with even higher pressures by shifting the W2-type Weyl points to the Fermi energy.

The pressure independence of the SdH frequencies evidences the robustness of the electronic structure of the WSM NbP in the pressure range up to 2.8~GPa. This observation is different compared with several 3D Dirac semimetals. For example, Cd$_{3}$As$_{2}$ exhibits a pressure-induced breakdown of the 3D Dirac semimetal state with the system becoming a gapped semiconductor at 2.57~GPa.\cite{S_Zhang} For WTe$_{2}$ a drastic change in the Fermi surface topology and a strong suppression of the MR was observed at 1.8~GPa.\cite{Cai} We speculate that the robustness of electronic structure of the WSM NbP may be related to its stable noncentrosymmetric tetragonal structure. This speculation is supported by recent studies on NbAs \cite{Luo2016NbAs,Zhang2015NbAs} and TaAs,\cite{Zhou2015TaAs} two others closely related members of the monopnictide family, with the same crystal structure than NbP. However, for both arsenide compounds,  the electrical resistivity displays at low temperatures a monotonic increase upon increasing pressure ($T\lesssim25$~K), in contrast to our finding of a pressure independent resistivity in NbP. This suggests that the electronic structure of NbP is more robust than that of NbAs and TaAs.

\section{SUMMARY}
In conclusion, we presented an investigation of the pressure evolution of the FS topology in the noncentrosymmetric Weyl semimetal NbP by combining experimental studies of SdH oscillations and band-structure calculations. We found a robust electronic structure in the pressure range up to 2.8 GPa. The characteristic topological features of the Fermi-surface with two electron and two hole pockets remain unchanged. The strong change in the SdH amplitudes with pressure can be attributed to subtle changes in the shape of the Fermi-surface in the vicinity of the extremal orbits. Furthermore, our results evidence a strong pressure variation of the magnitude of longitudinal magnetoresistance for $B\parallel a$  which might be a related to a combination of different effects, such as a change in the balance of electron and hole charge carriers and/or an alteration of the distance of the W2-type Weyl points to the Fermi level. 
Indeed, the W2-type Weyl points move toward the Fermi-energy upon increasing pressure, while the energy difference between the W1-type Weyl points and the Fermi energy is almost unaffected. We may speculate that the chiral anomaly effect might be established at even higher pressures. The good agreement between the experimental and calculated SdH frequencies confirm that the pressure dependence of the quantum oscillations provide a reliable tool to probe the Fermi-surface topology in Weyl semimetals under pressure.

\section*{ACKNOWLEDGMENTS}
We thank E. Hassinger, F. Arnold, and M. Baenitz for stimulating discussions. This work was financially supported by the ERC Advanced Grant No.\ (291472) ``Idea Heusler". R.\ dos Reis acknowledges financial support from the Brazilian agency CNPq.

\bibliography{NbP-references}

\end{document}